\documentclass[5p,sort&compress]{elsarticle}
\usepackage{amssymb}
\usepackage{amsmath,ragged2e}
\usepackage{color,appendix}
\usepackage{graphicx}
\usepackage{caption}
\usepackage{subcaption}
\usepackage[english]{babel}
\usepackage{bm}
\usepackage[percent]{overpic}
\usepackage{multirow,rotating}
\usepackage{epsfig}
\usepackage[english]{babel}

\newcommand{\er}{$\pm$}

\newcommand{\beq}{\begin{eqnarray}}
\newcommand{\eeq}{\end{eqnarray}}
\newcommand{\be}{\begin{eqnarray}}
\newcommand{\ee}{\end{eqnarray}}
\newcommand{\bea}{\begin{eqnarray}}
\newcommand{\eea}{\end{eqnarray}}

\definecolor{orange}{rgb}{1,.5,.3}
\definecolor{lightyellow}{cmyk}{0,0,0.5,0}
\definecolor{lightred}{rgb}{1,0.5,0.5}
\definecolor{lightgreen}{rgb}{0.8,1.0,0.8}
\definecolor{lightblue}{rgb}{0.5,0.5,1}
\definecolor{darkred}{rgb}{0.8,0,0}
\definecolor{darkgreen}{rgb}{0,0.4,0}
\definecolor{darkcyan}{cmyk}{1,0.3,0.3,0.3}
\definecolor{darkblue}{rgb}{0,0,0.6}
\definecolor{lightbrown}{rgb}{0.7,0.3,0.3}
\definecolor{darkbrown}{rgb}{0.5,0,0}
\definecolor{bluegreen}{rgb}{0,0.5,0.5}
\definecolor{mycyan}{rgb}{0.0,0.6,0.6}
\definecolor{mylila}{rgb}{0.6,0.,0.6}
\definecolor{mylila2}{rgb}{0.6,0.,0.5}
\definecolor{mygray}{rgb}{0.37,0.37,0.37}
\definecolor{myyellow2}{rgb}{0.8,0.5,0.1}
\definecolor{myyellow2b}{rgb}{0.9,0.9,0.1}
\definecolor{myyellow}{rgb}{0.6,0.6,0.}
\definecolor{mytuerkis}{rgb}{0.0,0.6,0.85}
\definecolor{grey0}{rgb}{0.7,0.7,0.7}
\definecolor{grey1}{rgb}{0.6,0.6,0.6}
\definecolor{grey2}{rgb}{0.4,0.4,0.4}
\definecolor{grey3}{rgb}{0.2,0.2,0.2}

\begin{document}
\begin{frontmatter}

\bibliographystyle{try}
\topmargin 0.1cm

\title{Search for the tensor glueball}

\author[label1]{E. Klempt}
\author[label2]{A.V. Sarantsev}
\author[label3]{I. Denisenko}
\author[label1]{and K.V. Nikonov}

\address[label1]{Helmholtz--Institut f\"ur Strahlen-- und Kernphysik, Universit\"at Bonn, Germany}
\address[label2]{NRC ``Kurchatov Institute'', PNPI, Gatchina 188300, Russia}
\address[label3]{Joint Institute for Nuclear Research, Joliot-Curie 6, 141980 Dubna, Moscow region, Russia}

\date{\today}
\begin{abstract}
The tensor glueball is searched for in BESIII data on radiative $J/\psi$ decays into $\pi^0\pi^0$
and $K_sK_s$.  The $\pi\pi$ invariant mass distribution exhibits an enhancement that can be 
described by a pole  at $(2210\pm 60) -i(180\pm 60)$\,MeV. We speculate if 
the tensor glueball could be distributed among high-mass tensor mesons.
\end{abstract}


\end{frontmatter}

\section{Introduction}

Quantum chromodynamics (QCD), the fundamental theory of strong interactions, predicts the
existence of a full spectrum of glueballs, of composite particles containing gluons but no valence quarks.
Their existence is a direct consequence of the nonabelian nature of QCD and of confinement.
The properties of glueballs have been studied in many models since their prediction in the 1970s
\cite{Fritzsch:1972jv,Fritzsch:1975tx} but experimentally, no generally accepted view had emerged.
Recent reviews of glueballs and of light-quark mesons can be found elsewhere
\cite{Klempt:2007cp,Mathieu:2008me,Crede:2008vw,Ochs:2013gi,Llanes-Estrada:2021evz}.
The scalar glueball is expected in the 1500 - 2000\,MeV mass 
range~\cite{Bali:1993fb,Morningstar:1999rf,Gregory:2012hu,%
Szczepaniak:2003mr,Rinaldi:2018yhf,Athenodorou:2020ani,Rinaldi:2021dxh,Chen:2021bck,%
Dudal:2021gif,Zhang:2021itx,Li:2021gsx}. Based on lattice calculations, we have to 
expect the tensor glueball mass 600 to 800\,MeV above the scalar glueball. 
However, a small mass gap is also possible. The authors of Ref.~\cite{Chen:2021bck} 
use QCD sum rules and predict 1780\,MeV for the scalar, and 1860\,MeV for the tensor glueball. 
  
The radiative decay
branching ratios for producing
glueballs were predicted by lattice gauge calculations \cite{Gui:2012gx,Chen:2014iua}:
\beq
\Gamma_{J/\psi\to\gamma/ G_{0^{++}}}/\Gamma_{\rm tot} &=&(3.8\pm0.9) 10^{-3}\,,\label{scalar-pred}\\
\Gamma_{J/\psi\to\gamma/ G_{2^{++}}}/\Gamma_{\rm tot} &=&(11\pm2) 10^{-3}\,.\label{tensor-pred}
\eeq
These are large numbers. The yield of $f_2(1270)$ in radiative $J/\psi$ decays is $(1.64\pm 0.12)10^{-3}$,
about six times weaker than the predicted rate for the tensor glueball!

Little is known about the glueball width. Arguments 
based on the $1/N_c$ expansion (see, e.g., Ref.~\cite{RuizdeElvira:2010cs}) suggest 
that glueballs might be narrow, 100\, MeV or less. Narison~\cite{Narison:1996fm}
gave a $2\pi$ partial decay width of the scalar glueball of $(119\pm36)$\,MeV, not incompatible 
with a large total width. Minkowski and Ochs assume a width exceeding 
1\,GeV~\cite{Minkowski:1998mf}. 
The authors of Refs.~\cite{Brunner:2015oqa,Brunner:2015yha} reproduce the decay 
rates of $f_0(1710)$ assuming that 
this is the scalar glueball and predict the tensor glueball to be very wide.

Recently we have presented a coupled-channel analysis~\cite{Sarantsev:2021ein} of
the $S$-wave amplitude from BESIII data on radiative $J/\psi$ decays. The data on
radiative $J/\psi$ into  $\pi^0\pi^0$ and $K_sK_s$
were published including a bin-wise partial-wave decomposition
into $S$-wave and $D$-wave ~\cite{Ablikim:2015umt,Ablikim:2018izx}, the data on decays into
for $\eta\eta$ and $\phi\omega$ were published only 
in an energy-dependent amplitude analysis~\cite{Ablikim:2013hq,Ablikim:2012ft}.
A large number of further data were included in the coupled-channel analysis, references
to the additional data can be found in Ref.~\cite{Sarantsev:2021ein}.
Ten scalar isoscalar resonances were required to fit the data. Five of them were interpreted
as mainly singlet, five as mainly octet resonances in SU(3). The yield of resonances
showed a striking peak at
$(1865\pm 25^{\,+10}_{\,-30})-i (185\pm 25^{\,+15}_{\,-10})$\,MeV
called $G_0(1865)$. In a subsequent paper we studied the decays of the scalar 
mesons into pairs of pseudoscalar mesons. We found that the decays can be understood
only when these mesons contain an additional flavor singlet fraction beyond the one
expected for any mixing angle.

This peak at 1865\,MeV showed properties expected from a scalar glueball: 
\begin{itemize}
\item $G_0(1865)$ is produced abundantly in radiative $J/\psi$ decays above a very
low background. Its mass is $1\sigma$ compatible with the mass calculated in unquenched 
lattice QCD~\cite{Gregory:2012hu} and the yield is $1.6\sigma$ compatible with the 
yield calculated in lattice QCD.\\[-4ex]
\item The decay analysis of the scalar isoscalar mesons shows that the assignment 
of mesons to mainly-octet and mainly-singlet states is correct. Even the production
of mainly-octet scalar mesons - which should be forbidden in radiative $J/\psi$ decays - 
peaks at 1865 MeV.\\[-4ex]
\item The decay analysis requires a small glueball content in the flavor wave
function of several scalar resonances.
The glueball content as a function of the mass shows a peak compatible with the peak in
the yield of scalar isoscalar mesons. The sum of the fractional glueball contributions is 
compatible with one~\cite{Klempt:2021wpg}.\\[-4ex]
\item In the reaction $B_s\to J/\psi + K^+K^-$, a primary $s\bar s$ couples to 
mesons having a strong coupling to $K^+K^-$~\cite{LHCb:2017hbp}.
Two peaks in the $K^+K^-$ mass spectrum are seen due to $\phi(1020)$ 
and $f_2'(1525)$ (see Fig.~7 in Ref.~\cite{LHCb:2017hbp}, Fig.~\ref{three}c below shows 
a fit to the tensor wave) but there is
no sign of higher-mass scalar mesons. In 
particular $f_0(1710)$ with its prominent $K\bar K$ decay mode is not seen in
$s\bar s\to  f_0(1710)\to K\bar K$. It is, however, produced very strongly in the process
gluon-gluon$\to  f_0(1710)\to K\bar K$: Obviously, 
$f_0(1710)$ is produced by two initial-state gluons but not by an $s\bar s$ pair in the initial state. 
$f_0(1710)$ must have a sizable glueball fraction! \\[-4ex]

\end{itemize}

For these reasons, we are convinced that the scalar glueball is distributed among the
agglomeration of scalar isoscalar mesons in the range from 1500 to 2300\,MeV. 
Based on lattice calculations, we have to expect the tensor glueball above 2500\,MeV 
even though smaller masses are possible as well~\cite{Chen:2021bck}. 
 
In this Letter we search for the tensor glueball expected to be produced in 
radiative $J/\psi$ decays into $\pi^0\pi^0$ and $K_sK_s$. In Section~\ref{II} 
we compare the intensities of the
$\pi\pi$ and $K\bar K$ invariant masses in the scalar and tensor wave. The tensor wave
reveals a wide high-mass resonance.
Subsequently, in Section~\ref{III}, we discuss
if the high-mass enhancement is split into several states and if these contain a small fraction 
of the tensor glueball.
The results are discussed and summarized in Section~\ref{IV}.

\section{\label{II}A high-mass tensor resonance from radiative $J/\psi$ decays
}
The $\pi^0\pi^0$ or $K_sK_s$ systems produced in radiative $J/\psi$ decays are
limited to even angular momenta due to Bose symmetry. Practically, only $S$ and $D$-waves
are relevant. These two partial waves can be written in the multipole basis~\cite{Rodas:2021tyb,Sebastian:1992xq}.
The scalar intensity originates from the electric dipole transition $E0$.
Three electromagnetic amplitudes, $E1, M2$, and $E3$, lead to the production of tensor mesons
where the $E1$ amplitude is the most significant one. These three amplitudes and relative phases
are discussed below.\vspace{-0mm}

\begin{figure}
\centering\vspace{-10mm}
\begin{overpic}[width=0.52\textwidth,height=0.6\textwidth]{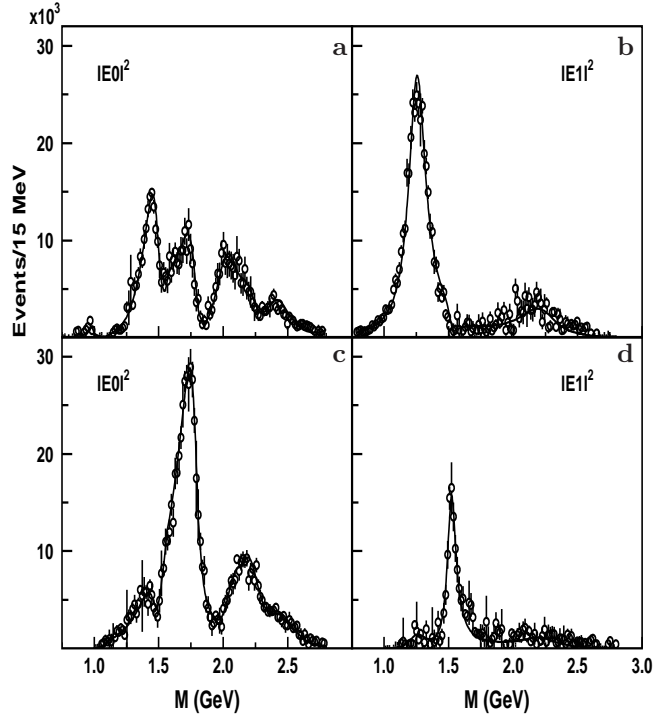}\vspace{-3mm}
\put(41,84){\bf a}
\put(75,84){\bf b}
\put(41,47){\bf c}
\put(75,47){\bf d}
\end{overpic}
\vspace{-6mm}
\caption{\label{one}The scalar (a,c) and tensor (b,d) intensities in radiative 
$J/\psi$ decays to $\pi^0\pi^0$
(a,b) and $K_sK_s$ (c,d) in 20\,MeV bins. The solid line is our fit. 
The data are from the BESIII collaboration \cite{Ablikim:2015umt,Ablikim:2013hq}.
}
\end{figure}

The $E0$ and $E1$ squared amplitudes lead to strikingly different mass distributions
(see Fig.~\ref{one}). The distributions were derived in Refs.
\cite{Ablikim:2015umt,Ablikim:2018izx} by exploiting the statistical precision 
provided by $(1.311\pm 0.011)\times 10^9 J/\psi$ decays collected by
the BESIII collaboration. 
The $\pi\pi$ mass distribution in the scalar partial wave (Fig.~\ref{one}a)
is characterized by a mountainous landscape starting with a steady rise growing up to 1450\,MeV and a
rapid fall-off. After a minimum, the intensity increases due to the $f_0(1710)/f_0(1770)$ complex.
After a further deep minimum, a wide and asymmetric structure due to $f_0(2020)/f_0(2100)$ follows.
At the highest mass, at about 2300\,MeV, a small dip-peak structure is seen.
In the $K\bar K$ distribution (Fig.~\ref{one}c), the intensity rises slightly up to 1450\,MeV, followed
by a very significant interference pattern and then rises steeply to an asymmetric peak
at about 1700\,MeV that dominates the mass distribution. After a fast drop on its right side,
a second peak at about 2100\,MeV appears with a high-mass shoulder.

The squared amplitudes that describe production of  tensor mesons do not show such rich structures.
Below 1\,GeV, the $\pi\pi-D$-wave (Fig. \ref{one}b) shows only a tail of the $f_2(1270)$
and no entries in $K\bar K$ mass distribution (Fig.~\ref{one}d). Above 1\,GeV the $\pi\pi$
$2^{++}$ intensity exhibits only one strong peak due to $f_2(1270)$ production and a wide
enhancement that reaches a maximum at about 2200\,MeV. The $K\bar K$ intensity
exhibits only one peak due to $f_2'(1525)$. Above, only little intensity is seen.

In a first fit, we describe the high-mass region by one additional
resonance.  Neither the mass distribution nor the phase difference
are well reproduced. The $\chi^2/N_{\rm data}= 1088/765$ for the mass distributions
and $2584/677$ for the phase differences. The mass distribution is reasonably described,
the phase differences qualitatively only. However,
apparent discrepancies are often enforced by adjacent high-statistics points, and some
structures are limited to a narrow mass window.

Alternatively, we allow for one further resonance
$f_2(1640)$, the fit improves only marginally. The fit with or without $f_2(1640)$
gives a narrower or wider high-mass tensor resonance. Since we do
not know if $f_2(1640)$ participates in the reaction, we increase
the errors correspondingly:
\be
M=(2210\pm 60)\,{\rm MeV}; \ \ \Gamma=(360\pm 120)\,{\rm
MeV}\,.
\ee
The error does not contain the possibility that the production amplitude of tensor mesons
may be reduced dynamically with decreasing photon energy. Only the phase space 
is taken into account. Tentatively, we call these resonances $X_{2}(2210)$
(not $f_2(2210)$ since it might be a cluster of resonances).

An energy-dependent partial-wave analysis of the same data on was reported
by the JPAC Collaboration~\cite{Rodas:2021tyb}.
Four scalar and three tensor mesons were identified.
The tensor amplitude was described by $f_2(1270)$, $f_2'(1525)$, and a further state at
about $f_2(1950)$ and a width of 700\,MeV. A possible tensor glueball was not discussed.
The difference in mass might be due a different choice of the ambiguous solutions of the
energy-independent partial wave analysis. With our choice, the $|E1|^2$ distribution
shows a clear peak at about 2200\,MeV. 

In our fit, $X_{2}(2210)$ was parameterized by a three-channel relativistic
Breit-Wigner amplitude with $\pi\pi$, $K\bar K$, and $\rho\rho$ as
decay channels. The ratio of the frequencies of $X_{2^{++}}(2210)$
decays into
$K\bar K$ and $\pi\pi$ is
\be
BR_{K\bar K/\pi\pi}=0.23\pm 0.05\,.
\ee
In Table~\ref{qqb} the properties  of $f_2(1270)$ and $f_2(1525)$ are compared with 
values given in the Review of Particle
Physics (RPP)~\cite{Zyla:2020zbs} and with other determinations using radiative $J/\psi$ decay.

The $f_2(1270)$ mass found here is incompatible with the RPP value. We note that in an analysis
of BESII data on $J/\psi\to \gamma\pi\pi$, the $f_2(1270)$ mass was determined to
$(1262^{+1}_{-2}\pm8)$\,MeV~\cite{Ablikim:2006db}, and from
CLEO data on this reaction, (1259\er4\er4)\,MeV was deduced~\cite{Dobbs:2015dwa}.
JPAC finds masses between 1262 and 1282\,MeV~\cite{Rodas:2021tyb}.

The $K\bar K/\pi\pi$ ratio for the $f_2(1270)$ could be determined with a large
uncertainty from the faint peak at about 1270\,MeV  in the $K\bar K$ mass distribution.
Here, we fix the ratio to the RPP value. Also, the small $\pi\pi$ decay mode of
$f_2'(1525)$ is fixed to the RPP value. Our radiative yields of $f_2(1270)$ and $f_2'(1525)$
yields are fully compatible with RPP values.

\begin{table}[h]
\caption{\label{qqb}Properties of $f_2(1270)$ and $f_2(1525)$. The RPP2021~\cite{Zyla:2020zbs}
values and those from Refs.~\cite{Ablikim:2006db,Dobbs:2015dwa} are given
in small numbers. Ratios marked (f) are fixed. The yields $Y$ of $f_2(1270)$ and $f_2(1525)$
are corrected for unseen decay modes;
the $X_2(2210)$ yield represents the sum of the $\pi\pi$ and $K\bar K$ yields. \vspace{-3mm}
}
\renewcommand{\arraystretch}{1.4}
\begin{center}
\begin{tabular}{ccccc}
\hline\hline
                          &          $f_2(1270)$              & $f_2'(1525)$ &           & $X_2(2210)$\\\hline
$M$  (MeV)          &          1257\er6                       & 1518\er3     &           &2210\er60 \\[-1.5ex]
                          &\scriptsize 1259\er6               & \scriptsize 1532\er7 &\scriptsize \cite{Dobbs:2015dwa}&\\[-1.5ex]
                         &\scriptsize 1275.5\er0.8           & \scriptsize 1517.4\er2.5 &\scriptsize \cite{Zyla:2020zbs}&\\[-1.5ex]
                          &\scriptsize 1262\er8               & \scriptsize  &\scriptsize \cite{Ablikim:2006db}&\\[-1.ex]
 $\Gamma$ (MeV)  &       168\er 7                              &          78\er6    &&  $360\pm120$                \\[-1.5ex]
                          &\scriptsize $185.9^{+2.8}_{-2.1}$  &  \scriptsize 86.9$^{2.3}_{2.1}$ &\scriptsize \cite{Zyla:2020zbs}  \\[-1.5ex]
                          &\scriptsize 175\er12               & \scriptsize  &\scriptsize \cite{Ablikim:2006db}&\\[-1.ex]
  $R_{K\bar K/\pi\pi}$   &                  0.054 (f)         &             && 0.23\er0.05                \\[-1.5ex]
                       &\scriptsize $0.054^{+0.005}_{-0.006}$ &            -                &\scriptsize \cite{Zyla:2020zbs}&    \\[-1ex]
$r_{\pi\pi/K\bar K}$   &                  -                                      &   0.0094 (f) &\scriptsize \cite{Zyla:2020zbs}               \\[-1.5ex]
                          &                 -                                      & \scriptsize 0.0094\er0.0018     \\[-1ex]
$Y (\times 10^3)$                   &            1.69\er0.07               &      0.61\er0.06 &&0.35\er0.10     \\[-1.5ex]
                       &\scriptsize  (2.08\er1.58)           & \scriptsize $<1.7$&\scriptsize \cite{Dobbs:2015dwa}\\[-1.5ex]
                        &\scriptsize  (1.63\er0.12)          & \scriptsize $(0.57^{+0.08}_{-0.05})$&\scriptsize \cite{Zyla:2020zbs}&\\[-1.5ex]
                         &\scriptsize  (1.63\er0.26)          & &\scriptsize \cite{Ablikim:2006db}\\
  \hline\hline
\end{tabular}
\end{center}
\end{table}

In the reaction $J/\psi\to\gamma$ plus a tensor meson, the production process couples to
the mesonic flavor-singlet component only. The ratio $R$ of $f_2'(1525)/f_2(1270)$ production
is related to the mixing angle via
\be
\tan^2\theta_{\rm tens}=\frac{1}{\lambda}\cdot R\cdot \frac{q_{ f_2}}{q_{f_2'}}
\ee
from which we find a tensor mixing angle $\theta^{\rm tens}=(35$$\pm$$2)^\circ$.
The mixing angle identifies the tensor meson nonet as {\it ideally mixed} but is
inconsistent with $29.8 (28.0)^\circ$ derived from the quadratic (linear) GMO formula.

\begin{figure*}
\centering\vspace{-5mm}
\begin{overpic}[width=0.95\textwidth,height=0.33\textwidth]{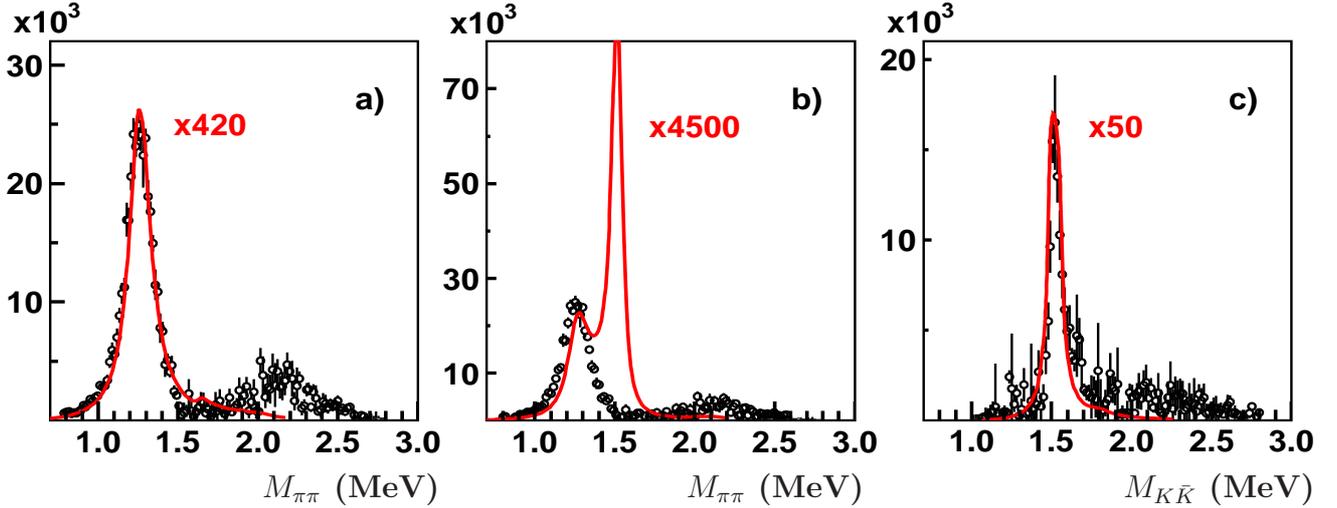}
\put(20,-3){\bf\large $M_{\pi\pi}$ (MeV)}
\put(52,-3){\bf\large $M_{\pi\pi}$ (MeV)}
\put(85,-3){\bf\large $M_{K\bar K}$ (MeV)}
\end{overpic}
\vspace{6mm}
\caption{\label{three}(Color online)The tensor intensities in radiative $J/\psi$ decays to $\pi^0\pi^0$
(a,b) and $K_sK_s$ (c) in 20\,MeV bins. The open circles correspond to the BESIII data
\cite{Ablikim:2015umt,Ablikim:2013hq}.
The solid (red) line represents the $\pi\pi$ mass distribution in the
tensor wave derived from data of the LHCb collaboration 
\cite{LHCb:2014vbo,LHCb:2014ooi,LHCb:2017hbp} on 
$B^0\to J/\psi + (\pi\pi)$ (a),  $B^0 _s\to J/\psi + (\pi\pi)$ (b), and on
$B^0 _s\to J/\psi + (K\bar K)$. The intensities from the  $B^0_{(s)}$ decays are
chosen to match the $f_2(1270)$ and $f_2'(1525)$ intensities from radiative $J/\psi$ decays.
}
\end{figure*}
We now need to ask: Is it plausible that just one tensor resonance above 1700\,MeV
is produced in radiative $J/\psi$ decays? There is at most marginal evidence for $f_{2}(1640)$, 
and no evidence at all for $f_{2}(1910/1950)$. Both states are seen in several 
experiments~\cite{Zyla:2020zbs}. 
But these states are at most very weakly produced in radiative $J/\psi$ decays. 
Above these two states, a tensor meson at 2210\,MeV suddenly appears.  
Could $X_{2}(2210)$ contain a fraction of the tensor glueball? And 
why is the fit to the
phases bad with a single-resonance fit? Are several tensor resonances hidden in $X_{2}(2210)$?

\section{\label{III}Could  $X_2(2210)$ be the tensor glueball\,?}

\subsection{The tensor wave in $B^0_{(s)}\to J/\psi + f_2$}
Figure~\ref{three} shows a comparison of the contribution of the $E1$ amplitude
in radiative $J/\psi$ decays and a fit to LHCb data on $B$ and $B_s$ decays. 
In Fig.~\ref{three}a,b the data from Fig.~\ref{one}b are reproduced, in 
Fig.~\ref{three}c the data from Fig.~\ref{one}d. Superimposed is a fit 
to the $\pi\pi$ and $K\bar K$ $D$-wave contributions to 
$B^0\to J/\psi + (\pi\pi)$~\cite{LHCb:2014vbo} (a), 
$B^0 _s\to J/\psi +(\pi\pi)$~\cite{LHCb:2014ooi} (b), and
$B^0 _s\to J/\psi +(K\bar K)$~\cite{LHCb:2017hbp} (c). 
In $B^0\to J/\psi$ $+$ hadrons, the $b$ quark converts into a $c$ quark
radiating off a $W^-$ boson. The $W^-$ boson decays into a $\bar c$ plus a $d$ quark. 
The $c\bar c$ is seen as $J/\psi$, the $d\bar d$ forms a light-quark meson. 
In $B^0 _s\to J/\psi$ $+$ hadrons, the $W^-$ boson decays into a $\bar c$ plus a $s$ quark,
and an $s\bar s$ pair creates the final state.
The phase space of the BESIII data on radiative $J/\psi$ extends up to 3.1\,GeV,
the phase space in the reaction $B^0\to J/\psi$ $+$ hadrons is limited to 2180\,MeV,
and to 2270\,MeV in the case of $B_s\to J/\psi$ $+$ hadrons.

The main objective
of the LHCb collaboration was the study of CP violation through the interference
of $B^0 _{(s)}\leftrightarrow\bar B^0_{(s)}$ and their decay amplitudes. But the resonant 
structures in the $\pi\pi$ and $K\bar K$ system were studied as well. The
angular distributions were presented in the form of spherical harmonic moments.
We have included these spherical harmonic moments 
into the data set described above for a joint coupled-channel analysis. The
main results are presented elsewhere~\cite{Sarantsev:2022tbd}. The fit returns
the  $\pi\pi$ and $K\bar K$ $S$, $P$ and $D$-wave amplitudes recoiling 
against the $J/\psi$. 

The $\pi\pi$ $D$-wave from LHCb in Fig.~\ref{three}a shows a peak due 
to $f_2(1270)$ production, the $K\bar K$ $D$-wave Fig.~\ref{three}c
a peak due to $f_2'(1525)$. The intensities were multiplied by factors
 given in the subfigures. The factors are chosen to match the $f_2(1270)$ or $f_2'(1525)$ peak
heights of the results from radiative $J/\psi$ decays. Masses and
widths are well compatible. The solid curve in Fig.~\ref{three}b  exhibits
a double-peak structure; both, $f_2(1270)$ and $f_2'(1525)$, contribute
to this reaction.

First, we discuss the overall intensities of the LHCb data. The strongest reaction,
$B^0 _s\to J/\psi +f_2'(1525)$, $f_2'(1525)$ $\to K\bar K$, is about 12 times
stronger than $B^0 \to J/\psi +f_2(1270),$ $f_2(1270)\to \pi\pi$. In the former
reaction, the intermediate $W$ boson converts into a $c$ and an $s$
quark, in the latter reaction into a $c$ and a $d$ quark. From the ratio of the 
CKM matrix elements $|V_{cd}/V_{cs}|^2$, we expect a large reduction.
 
There is very weak intensity only in the LHCb data above the $f_2(1270)$ or $f_2'(1525)$.
If the peak at 
2210\,MeV in Fig.~\ref{one}b were due to a regular $q\bar q$ state, we would expect
an onset of the tensor intensity in the LHCb data, in particular
in Fig.~\ref{three}a. This is not the case. As in the case
of scalar mesons, the high-mass enhancement is not produced by $q\bar q$ 
in the initial state but by gluon-gluon interactions. The enhancement at 2210\,MeV
seems to contain a significant fraction of the tensor glueball in its wave function. 
Due to the limited phase space, this argument is, however, suggestive only and not really
enforcing.

\subsection{$\phi\phi$ decays of tensor mesons}
The scalar glueball was distributed among several scalar isoscalar resonances.
Hence we expect that also the tensor glueball might not be concentrated
in a single resonance. 
Etkin {\it et al.}~\cite{Etkin:1987rj} at BNL observed a strikingly high intensity
above 2000\,MeV in the reaction $\pi^-p\to \phi\phi n$. The intensity was fully
ascribed to the $J^{PC}=2^{++}$ wave and was described by three tensor resonances
with masses and widths of about 
$(M,\Gamma) = (2010, 200)$\,MeV,  $(2300, 150)$\,MeV,  and
$(2340, 320)$\,MeV.  
The unusual production characteristics were interpreted in Ref.~\cite{Etkin:1987rj}
as evidence that {\it these states are produced by $1 - 3$ glueballs}.
The BESIII collaboration studied the process $J/\psi\to\gamma\phi\phi$ and found 
that the tensor wave of this reaction can be described well with these three tensor 
mesons \cite{BESIII:2016qzq}.  
The mean mass of the three $\phi\phi$ resonances is
2215\,MeV. This mass agrees perfectly well with the mass of $X_{2^{++}}(2210)$. 
The BNL experiment may thus have revealed 
the tensor glueball and its splitting into
several tensor mesons 40 years ago~\cite{Etkin:1982bw}!

\subsection{Fits with $X_2(2210)$ as cluster of resonances}  
\begin{figure*}[pt]
\begin{center}
\begin{tabular}{cc}
\begin{overpic}[width=0.5\textwidth,height=0.62\textwidth]{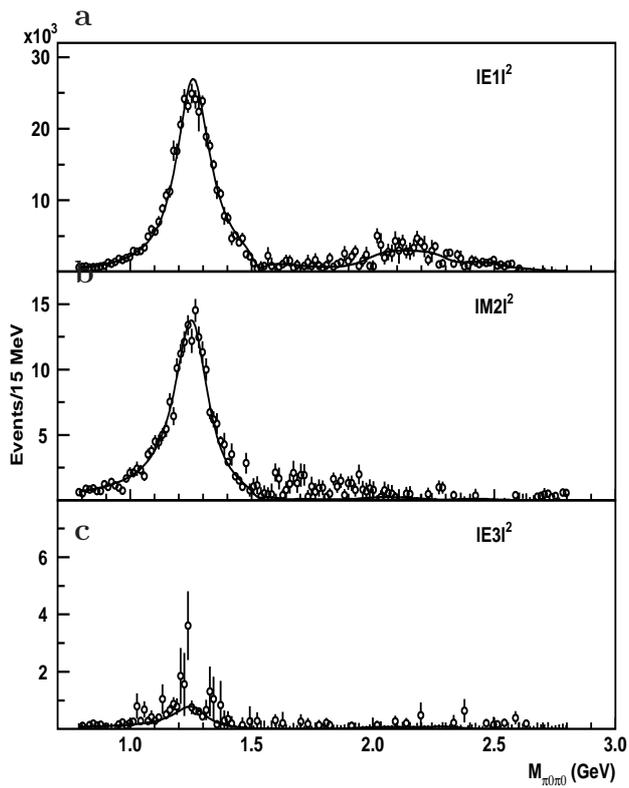}
\put(10,92){\bf\large a}
\put(10,62){\bf\large b}
\put(10,32){\bf\large c}
\end{overpic}&\hspace{-6mm}
\begin{overpic}[width=0.5\textwidth,height=0.62\textwidth]{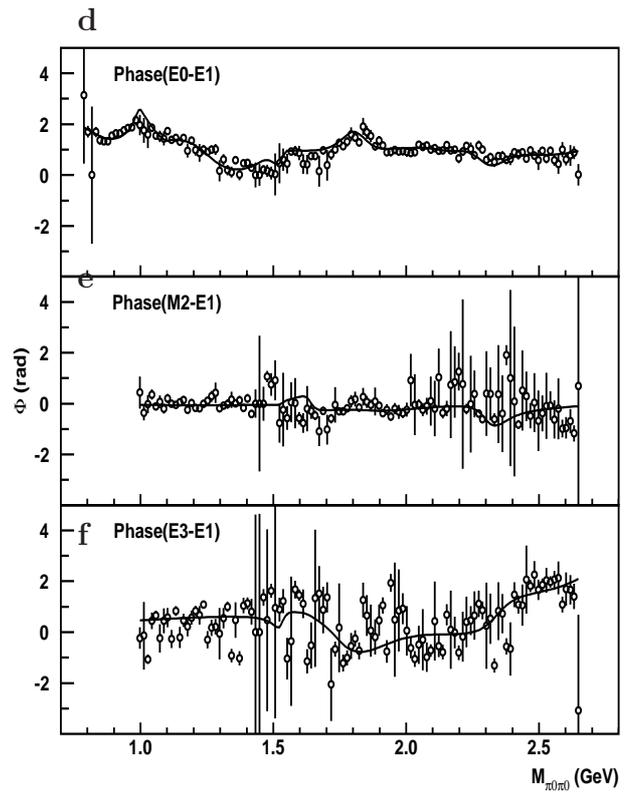}
\put(10,92){\bf\large d}
\put(10,62){\bf\large e}
\put(10,32){\bf\large f}
\end{overpic}\\[-2ex]
\begin{overpic}[width=0.5\textwidth,height=0.62\textwidth]{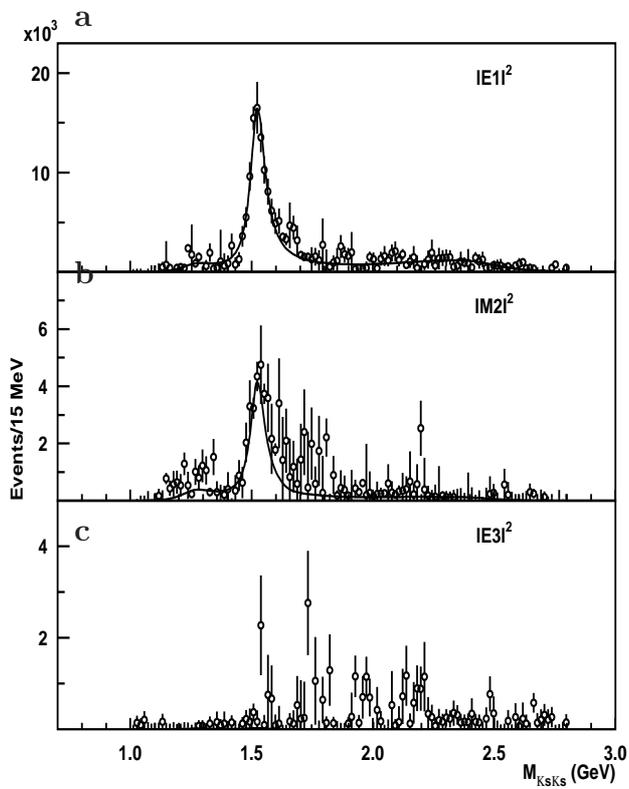}
\put(10,92){\bf\large a}
\put(10,62){\bf\large b}
\put(10,32){\bf\large c}
\end{overpic}&\hspace{-6mm}
\begin{overpic}[width=0.5\textwidth,height=0.62\textwidth]{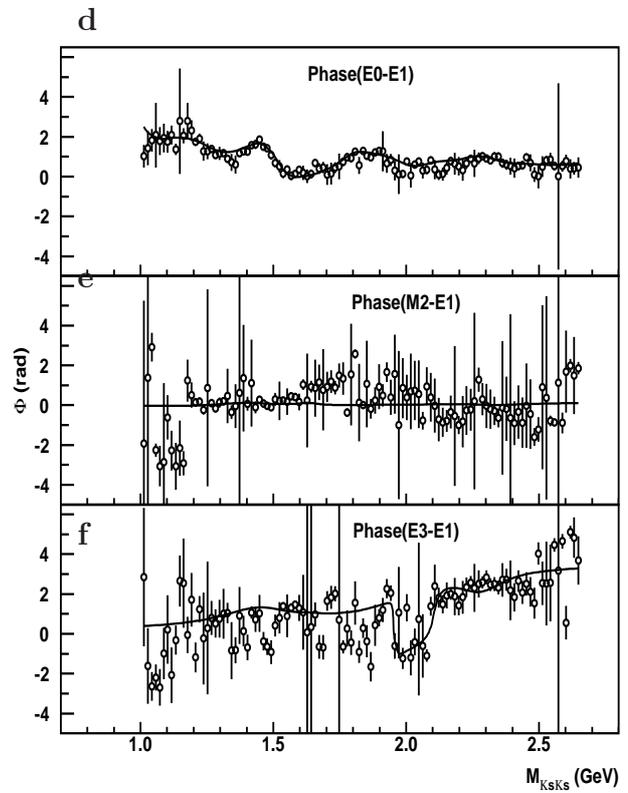}
\put(10,92){\bf\large d}
\put(10,62){\bf\large e}
\put(10,32){\bf\large f}
\end{overpic}
\end{tabular}
\vspace{-5mm}\end{center}
\caption{\label{jpsia}$D$-wave intensities and phases for radiative $J/\psi$ decays into 
$\pi^0\pi^0$ (top subfigures) and $K_s\,K_s$ (bottom subfigures)  from
Ref.~\cite{Ablikim:2015umt,Ablikim:2018izx}. The subfigures show 
the $E1$ (a), $M2$ (b) and $E3$ (c) squared amplitudes 
and the phase differences between the $E0$ and $E1$ (d) amplitudes, the $M2$ and $E1$ (e) amplitudes,
and the $E3$ and $E1$ (f) amplitudes as functions of the meson-meson invariant mass. The
phase of the $E0$ amplitude is set to zero. The  curve represents our best fit.
}
\end{figure*}

We fitted high-mass enhancement at 2210\,MeV
with these three resonances. Figure~\ref{jpsia} shows for the reaction  
$J/\psi\to \gamma \pi^0\pi^0$ and $K_sK_s$
\begin{enumerate}
\item the magnitudes of the three amplitudes $E1; M2$, and $E3$, and
\item the phase difference between the $E0$ and $E1$, $M2$ and $E1$, $E3$ and $E1$
amplitudes.
\end{enumerate}

The amplitudes - moduli and phases - are shown here in a mass region
limited to $0.75-2.75$\,GeV. The amplitudes were determined in slices of
the invariant mass in a “mass-independent fit”.  It was
shown that at each invariant mass, two solutions exist. Assuming
continuity of the amplitude, the full mass range could be described
by four different solutions. One of the solutions gave the best
energy-dependent fit for the scalar wave~\cite{Sarantsev:2021ein}.
This solution also defines unambiguously the tensor waves. 

The tensor intensities $E1$, $M2$, and $E3$ and the phase differences of
this solution are shown as histograms. The solid curve represents
our fit to the $S$ and $D$-waves. The $S$-wave was refit; the changes
of $S$-wave parameters compared to Ref.~\cite{Sarantsev:2021ein} are
marginal only. The $\chi^2$ of the overall fit  
is now $\chi^2/N_{\rm data}=890/765$ for the mass distributions and 
1716/677 for the phase differences. Adding further high-mass tensor resonances
improves the fit only slightly.

A few regions need to be discussed. 
The $M2$ yields for $\pi^0\pi^0$ and $K_sK_s$ yields above the $f_2'(1525)$
are underestimated by our fit. Larger yields are, however, incompatible with the phase motions.
The $M2-E1$ and $E3-E1$ phases have data points with very small errors and large
deviations from their neighbors. The most important phase difference $E0-E1$ is described by 
$\chi^2/N_{\rm data}=298/245$. 

\section{\label{IV}Discussion and Summary}

The total observed yield in $\pi\pi$ and $K\bar K$ is
\be
\hspace{-3mm}\sum_{M=1.9\,{\rm GeV}}^{M=2.5\,{\rm GeV}} Y_{J/\psi \to \gamma f_2, f_2\to \pi\pi, K\bar K} = (0.35\pm 0.15)\,10^{-3}\,.
\ee
Data on $\pi\pi$ elastic $D$-wave scattering in this mass range do not exist. The missing intensity
cannot be determined from the data included in our fits. An estimate can be obtained from
tensor states reported to be seen in radiative $J/\psi$ decays. The reactions and their
contributions to the high-mass region are listed in Table~\ref{sumyields}.
Summation yields
\be
\sum_{M=1.9\,{\rm GeV}}^{M=2.5\,{\rm GeV}} Y_{J/\psi \to \gamma f_2} = (3.1\pm 0.6)\,10^{-3}\,.
\ee
This is a substantial yield even though still smaller than the observed yield of the scalar glueball.
We note that the $4\pi$ tensor contribution is rather small when compared to
the $K^*(892)\bar K^*(892)$ and $\phi\phi$ contributions. The $6\pi$ tensor contribution
is completely unknown. Hence there may still be missing intensity. Here we emphasize
the importance of further studies of
these channels with the much larger statistics taken by the BESIII Collaboration.

\begin{table}
\caption{\label{sumyields}Yield of tensor mesons above 1900\,MeV 
in radiative $J/\psi$ decays in units 
of $10^{-5}$. The $4\pi$ yield is calculated from the $J/\psi\to\pi^+\pi^-\pi^+\pi^-$
yield by multiplication with the factor 9/4. The sum of all measured yields is  
$(3.1\pm0.6)\cdot 10^{-3}$.
}
\renewcommand{\arraystretch}{1.3}
\centering
\begin{tabular}{ccc}
\hline\hline
$\pi\pi$ &   $f_0(2210)$ & This work\\[-1.6ex]
                &  \tiny $(35\pm4)$ & \\[-0.5ex]\hline
$K\bar K$ & $f_0(2210)$ & This work\\[-1.6ex]
                &  \tiny $6\pm3$ &   \\[-0.5ex]\hline
$\eta\eta$ & $f_2(2340)$ &\cite{Ablikim:2013hq}\\[-1.2ex]
                &  \tiny  $(5.6^{+2.5}_{-2.2 })$ & \\[-0.5ex]\hline
$\eta\eta'$ &$f_2(2010)$, $f_2(2340)$& \cite{BESIII:2022qzu}\\[-1.6ex]
                &  \tiny $(1.36\pm0.10)$ $(0.25\pm0.04)$& \\[-0.5ex]\hline
$\eta'\eta'$ & $f_2(2340)$ &\cite{BESIII:2022zel}\\[-1.2ex]
                &  \tiny $(8.7^{+0.9}_{-1.8})$& \\[-0.5ex]\hline
$4\pi$ & $f_2(1950)$ & \cite{BES:1999dmf} \\[-1.6ex]
                &  \tiny  (124\er 43) & \\[-0.5ex]\hline
$\omega\omega$ &$f_2(1910)$ &\cite{BES:2006nqh}\\[-1.6ex]
                &  \tiny  (28\er 18)& \\[-0.5ex]\hline
$K^*\bar{K} ^*$ &$f_2(1950)$  &\cite{BES:1999zaa}\\[-1.6ex]
                &  \tiny  (70\er 23) &   \\[-0.5ex]\hline
$\phi\phi$ &$f_2(2010)$, $f_2(2300)$, $f_2(2340)$ &\cite{BESIII:2016qzq}\\[-1.6ex]
                &  \tiny $(3.5^{+3.2}_{-1.6 })$,    $(4.4^{+1.1}_{-1.7 })$,  $(19.1^{+7.3}_{-7.4 })$ & \\
\hline\hline
\end{tabular}
\end{table}

Summarizing, we have presented a coupled-channel analysis of BESIII data on
$J/\psi$ decays into $\pi^0\pi^0$ and $K_sK_s$. The data are dominated by $S$-wave and
$D$-wave contributions. This fit is important since it 
does not only provide the tensor wave but 
also shows that $S$ and $D$-waves both are consistently described. 

In the tensor wave we find an enhancement at $M=(2210\pm 60)$\,MeV, 
$\Gamma=360\pm120$\,MeV
and a yield (in $\pi\pi$ and $K\bar K$) of $(0.35\pm 0.10)\,10^{-3}$
called $X_2(2210)$.
There are arguments speaking in favor and against a glueball interpretation.\\[-5ex]
\paragraph{Pro:}   $X_2(2210)$ is produced as a high-mass tensor resonance
in radiative $J/\psi$ decays, a process in which glueballs are supposed to be produced.
It is the only tensor meson seen clearly above $f_2'(1525)$. This suggests that $X_2(2210)$ could
contain a contribution from the tensor glueball.
The improvement of the fit with three resonances instead of one only points to the
possibility that $X_2(2210)$ is composed of several resonances. In particular,
$X_2(2210)$ is consistent with early results in $\phi\phi$ production in $\pi N$ scattering
and supports the claim that this could be the tensor glueball.
 
There is no evidence for a similar enhancement in the reactions $B^0\to J/\psi + (\pi\pi)$ 
or  $B^0_s\to J/\psi + (\pi\pi)$ studied by the LHCb collaboration where resonances 
in the $(\pi\pi)$ are formed by a $d\bar d$ pair in the initial state. The absence of a structure
can serve as additional evidence for the glueball interpretation.\\[-5ex]
\paragraph{Contra:} In calculation on a lattice, mass and yield of the scalar glueball 
are predicted which agree well with the result of a coupled-channel analysis of the
same data as discussed here. These calculations predict a tensor glueball mass
considerably above $X_2(2210)$. Also the yield of the tensor glueball should be
substantially larger than the $X_2(2210)$ yield. The LHCb data are limited in phase
space, and the non-observation of a signal could be a phase-space effect. \\[-3ex]

Further studies of radiative $J/\psi$ decays are certainly required to support or
to reject the possibility that $X_2(2210)$ is the tensor glueball. Possibly, $X_2(2210)$
is only the low-energy tail of a tensor glueball centered at a higher mass. This conjecture
could be tested by analyzing data on radiative $\psi(2S)$ decays.

\section*{Acknowledgement}
Funded by the Deutsche Forschungsgemeinschaft (DFG, German Research
Foundation) – Project-ID 196253076 – TRR 110T and 
the Russian Science Foundation (RSF 22-22-00722).

\boldmath

\begin{thebibliography}{99}
%
\bibitem{Fritzsch:1972jv}
H.~Fritzsch and M.~Gell-Mann,
``Current algebra: Quarks and what else?,''
eConf \textbf{C720906V2}, 135 (1972).
%
\bibitem{Fritzsch:1975tx}
H.~Fritzsch and P.~Minkowski,
``Psi Resonances, Gluons and the Zweig Rule,''
Nuovo Cim. A \textbf{30}, 393 (1975).
%
\bibitem{Klempt:2007cp}
  E.~Klempt and A.~Zaitsev,
  ``Glueballs, Hybrids, Multiquarks. Experimental facts versus QCD inspired concepts,''
  Phys.\ Rept.\  {\bf 454}, 1 (2007).
%
\bibitem{Mathieu:2008me}
V.~Mathieu, N.~Kochelev and V.~Vento,
``The Physics of Glueballs,''
Int. J. Mod. Phys. E \textbf{18}, 1 (2009).
%
\bibitem{Crede:2008vw}
V.~Crede and C.~A.~Meyer,
``The Experimental Status of Glueballs,''
Prog. Part. Nucl. Phys. \textbf{63}, 74 (2009).
%
\bibitem{Ochs:2013gi}
W.~Ochs,
``The Status of Glueballs,''
J. Phys. G \textbf{40}, 043001 (2013).
%
\bibitem{Llanes-Estrada:2021evz}
F.~J.~Llanes-Estrada,
``Glueballs as the Ithaca of meson spectroscopy,''
Eur. Phys. J. ST \textbf{230}, no.6, 1575-1592 (2021).
%
\bibitem{Bali:1993fb}
G.~S.~Bali \textit{et al.} [UKQCD Collaboration],
``A Comprehensive lattice study of SU(3) glueballs,''
Phys. Lett. B \textbf{309}, 378 (1993).
%
\bibitem{Morningstar:1999rf}
  C.~J.~Morningstar and M.~J.~Peardon,
  ``The Glueball spectrum from an anisotropic lattice study,''
  Phys.\ Rev.\ D {\bf 60}, 034509 (1999).
 %
\bibitem{Gregory:2012hu}
  E.~Gregory, A.~Irving, B.~Lucini, C.~McNeile, A.~Rago, C.~Richards and E.~Rinaldi,
  ``Towards the glueball spectrum from unquenched lattice QCD,''
  JHEP {\bf 1210}, 170 (2012).
%
\bibitem{Szczepaniak:2003mr}
A.~P.~Szczepaniak and E.~S.~Swanson,
``The low lying glueball spectrum,''
Phys. Lett. B \textbf{577}, 61-66 (2003).
%
\bibitem{Rinaldi:2018yhf}
M.~Rinaldi and V.~Vento,
``Pure glueball states in a Light-Front holographic approach,''
J. Phys. G \textbf{47}, no.5, 055104 (2020).
%
\bibitem{Athenodorou:2020ani}
A.~Athenodorou and M.~Teper,
``The glueball spectrum of SU(3) gauge theory in 3 + 1 dimensions,''
JHEP \textbf{11}, 172 (2020).
%
\bibitem{Rinaldi:2021dxh}
M.~Rinaldi and V.~Vento,
``Meson and glueball spectroscopy within the graviton soft wall model,''
Phys. Rev. D \textbf{104}, no.3, 034016 (2021).
%
\bibitem{Chen:2021bck}
H.~X.~Chen, W.~Chen and S.~L.~Zhu,
``Two- and three-gluon glueballs of $C=+$,''
[arXiv:2107.05271 [hep-ph]].
%
\bibitem{Dudal:2021gif}
D.~Dudal, O.~Oliveira and M.~Roelfs,
``K$\ddot{a}$ll$\acute{e}$n-Lehmann Spectral Representation of the Scalar SU(2) Glueball,''
[arXiv:2103.11846 [hep-lat]].
%
\bibitem{Zhang:2021itx}
L.~Zhang, C.~Chen, Y.~Chen and M.~Huang,
``Spectra of glueballs and oddballs and the equation of state from holographic QCD,''
[arXiv:2106.10748 [hep-ph]].
%
\bibitem{Li:2021gsx}
H.~n.~Li,
``Dispersive analysis of glueball masses,''
Phys. Rev. D \textbf{104}, no.11, 114017 (2021).
%
\bibitem{RuizdeElvira:2010cs}
J.~Ruiz de Elvira, J.~R.~Pelaez, M.~R.~Pennington and D.~J.~Wilson,
``Chiral Perturbation Theory, the ${1/N_c}$ expansion and Regge behaviour 
determine the structure of the lightest scalar meson,''
Phys. Rev. D \textbf{84}, 096006 (2011).
%
\bibitem{Narison:1996fm}
S.~Narison,
``Masses, decays and mixings of gluonia in QCD,''
Nucl. Phys. B \textbf{509}, 312-356 (1998).
%
\bibitem{Minkowski:1998mf}
P.~Minkowski and W.~Ochs,
``Identification of the glueballs and the scalar meson nonet of lowest mass,''
Eur. Phys. J. C \textbf{9}, 283-312 (1999).
%
\bibitem{Brunner:2015oqa}
F.~Br\"unner, D.~Parganlija and A.~Rebhan,
``Glueball Decay Rates in the Witten-Sakai-Sugimoto Model,''
Phys. Rev. D \textbf{91}, no.10, 106002 (2015)
[erratum: Phys. Rev. D \textbf{93}, no.10, 109903 (2016)].
%
\bibitem{Brunner:2015yha}
F.~Br\"unner and A.~Rebhan,
``Nonchiral enhancement of scalar glueball decay in the Witten-Sakai-Sugimoto model,''
Phys. Rev. Lett. \textbf{115}, no.13, 131601 (2015).
%
\bibitem{Gui:2012gx}
L.~C.~Gui \textit{et al.} [CLQCD],
``Scalar Glueball in Radiative $J/\psi$ Decay on the Lattice,''
Phys. Rev. Lett. \textbf{110} no.2, 021601 (2013).
%
\bibitem{Chen:2014iua}
Y.~Chen {\it et al.},
``Glueballs in charmonia radiative decays,''
PoS \textbf{LATTICE2013}, 435 (2014).
%
\bibitem{Sarantsev:2021ein}
A.V. Sarantsev, I. Denisenko, U. Thoma, E. Klempt,
``Scalar isoscalar mesons and the scalar glueball from radiative $J/\psi$ decay,"
Phys. Lett. B \textbf{816}, 136227 (2021).
%
\bibitem{Ablikim:2015umt}
M.~Ablikim \textit{et al.} [BESIII Collaboration],
``Amplitude analysis of the $\pi^0\pi^0$ system produced in radiative $J/\psi$ decays,''
Phys. Rev. D \textbf{92} no.5, 052003 (2015).
%
\bibitem{Ablikim:2018izx}
M.~Ablikim \textit{et al.} [BESIII Collaboration],
``Amplitude analysis of the $K_{S}K_{S}$ system produced in radiative $J/\psi$ decays,''
Phys. Rev. D \textbf{98}  no.7, 072003 (2018).
%
\bibitem{Ablikim:2013hq}
  M.~Ablikim {\it et al.} [BESIII Collaboration],
  ``Partial wave analysis of $J/\psi \to \gamma \eta \eta$,''
  Phys.\ Rev.\ D {\bf 87}, no. 9, 092009 (2013).
%
\bibitem{Ablikim:2012ft}
M.~Ablikim \textit{et al.} [[BESIII Collaboration],
``Study of the near-threshold $\omega\phi$ mass enhancement in doubly
OZI-suppressed $J/\psi\rightarrow\gamma\omega\phi$ decays,''
Phys. Rev. D \textbf{87} no.3, 032008 (2013).
%
\bibitem{Klempt:2021wpg}
E.~Klempt and A.~V.~Sarantsev,
``Singlet-octet-glueball mixing of scalar mesons,''
Phys. Lett. B \textbf{826}, 136906 (2022).
%
\bibitem{LHCb:2017hbp}
R.~Aaij \textit{et al.} [LHCb],
``Resonances and $CP$ violation in $B_s^0$ and $\overline{B}_s^0 \to J/\psi K^+K^-$ decays in the mass region above the $\phi(1020)$,''
JHEP \textbf{08}, 037 (2017).
%
\bibitem{Sebastian:1992xq}
K.~J.~Sebastian, H.~Grotch and F.~L.~Ridener,
``Multipole amplitudes in parity changing one photon transitions of charmonium,''
Phys. Rev. D \textbf{45}, 3163-3172 (1992).
%
\bibitem{Rodas:2021tyb}
A.~Rodas \textit{et al.} [JPAC],
``Scalar and tensor resonances in radiative $J/\psi$ decays,''
Eur. Phys. J. C \textbf{82}, no.1, 80 (2022).
%
\bibitem{Zyla:2020zbs}
P.~A.~Zyla \textit{et al.} [Particle Data Group],
``Review of Particle Physics,''
PTEP \textbf{2020}, no.8, 083C01 (2020).
%
\bibitem{Dobbs:2015dwa}
S.~Dobbs, A.~Tomaradze, T.~Xiao and K.~K.~Seth,
``Comprehensive Study of the Radiative Decays of $J/\psi$ and $\psi(2S)$ to Pseudoscalar Meson Pairs, and Search for Glueballs,''
Phys. Rev. D \textbf{91}, no.5, 052006 (2015).
%
\bibitem{Zyla:2020zbs}
P.~A.~Zyla \textit{et al.} [Particle Data Group],
``Review of Particle Physics,''
PTEP \textbf{2020}, no.8, 083C01 (2020).
%
\bibitem{Ablikim:2006db}
M.~Ablikim \textit{et al.} [BES], 
``Partial wave analyses of $J/\psi \to \gamma \pi^+ \pi^-$ and $\gamma \pi^0 \pi^0$,''
Phys. Lett. B \textbf{642}, 441-448 (2006).
%
\bibitem{LHCb:2014vbo}
R.~Aaij \textit{et al.} [LHCb],
``Measurement of the resonant and CP components in $\overline{B}^0\to J/\psi \pi^+\pi^-$ decays,''
Phys. Rev. D \textbf{90}, no.1, 012003 (2014).
%
\bibitem{LHCb:2014ooi}
R.~Aaij \textit{et al.} [LHCb],
``Measurement of resonant and CP components in $\bar{B}_s^0\to J/\psi\pi^+\pi^-$ decays,''
Phys. Rev. D \textbf{89}, no.9, 092006 (2014).
%
\bibitem{Sarantsev:2022tbd}
A.~V.~Sarantsev {\it et al.}, 
``Scalar and tensor mesons in $d\bar d$, $s\bar s$ and  $gg\to f_0$," 
in preparation. 
%
\bibitem{Etkin:1987rj}
A.~Etkin \textit{et al.},
``Increased Statistics and Observation of the $g(T$), $g(T$)-prime, and $g(T$)-prime-prime $2^{++}$ Resonances in the Glueball Enhanced Channel $\pi^- p \to \phi \phi n$,''
Phys. Lett. B \textbf{201}, 568-572 (1988).
%
\bibitem{BESIII:2016qzq}
M.~Ablikim \textit{et al.} [BESIII],
``Observation of pseudoscalar and tensor resonances in $J/\psi\to \gamma \phi \phi$,''
Phys. Rev. D \textbf{93}, no.11, 112011 (2016).
%
\bibitem{Etkin:1982bw}
A.~Etkin \textit{et al.},
``The Reaction $\pi^- p \to \phi \phi n$ and Evidence for Glueballs,''
Phys. Rev. Lett. \textbf{49}, 1620 (1982).
%
\bibitem{BESIII:2022qzu}
M.~Ablikim \textit{et al.} [BESIII],
``Partial wave analysis of $J/\psi\rightarrow\gamma\eta\eta'$,''
[arXiv:2202.00623 [hep-ex]].
%
\bibitem{BESIII:2022zel}
M.~Ablikim \textit{et al.} [BESIII],
``Partial wave analysis of $J/\psi \to \gamma \eta^{\prime} \eta^{\prime}$,''
[arXiv:2201.09710 [hep-ex]].
%
\bibitem{BES:1999dmf}
J.~Z.~Bai \textit{et al.} [BES],
``Partial wave analysis of $J/\psi \to \gamma (\pi^+\pi^-\pi^+\pi^-)$,''
Phys. Lett. B \textbf{472}, 207-214 (2000).
%
\bibitem{BES:2006nqh}
M.~Ablikim \textit{et al.} [BES],
``Pseudoscalar production at $\omega\omega$ threshold in $J/\psi\to\gamma\omega\omega$,''
Phys. Rev. D \textbf{73}, 112007 (2006).
%
\bibitem{BES:1999zaa}
J.~Z.~Bai \textit{et al.} [BES],
``Partial wave analysis of $J/\psi \to \gamma (K^+K^-\pi^+\pi^-)$,''
Phys. Lett. B \textbf{472}, 200-206 (2000).
%
 \bibitem{Ablikim:2006ca}
  M.~Ablikim {\it et al.} [BES Collaboration],
  ``Pseudoscalar production at $\omega\omega$ threshold in $J/\psi\to \gamma \omega \omega$,''
  Phys.\ Rev.\ D {\bf 73}, 112007 (2006).
%
\bibitem{Alde:1998mc}
D.~Alde \textit{et al.} [GAMS Collaboration],
``Study of the $\pi^0\pi^0$ system with the GAMS-4000 spectrometer at 100\,GeV/c,''
Eur. Phys. J. A \textbf{3}, 361 (1998).
%
\bibitem{Longacre:1986fh}
R.~S.~Longacre \textit{et al.},
``A Measurement of $\pi^- p \to K_S K_S n$ at 22\,GeV/c and a Systematic Study of the $2^{++}$ Meson Spectrum,''
Phys. Lett. B \textbf{177}, 223 (1986).
%
\bibitem{Lindenbaum:1991tq}
S.~J.~Lindenbaum and R.~S.~Longacre,
``Coupled channel analysis of $J^{PC} = 0^{++}$ and $2^{++}$ isoscalar mesons with masses below 2\,GeV,''
Phys. Lett. B \textbf{274}, 492 (1992).
%
\bibitem{Grayer:1974cr}
G.~Grayer \textit{et al.}, 
``High Statistics Study of the Reaction $\pi^- p \to \pi^- \pi^+ n$: Apparatus, Method of Analysis, and General Features of Results at 17\,GeV/c,''
Nucl. Phys. B \textbf{75}, 189 (1974).
%
\bibitem{Batley:2010zza}
J.~R.~Batley \textit{et al.} [NA48/2 Collaboration],
``Precise tests of low energy QCD from $K_{e4}$ decay properties,''
Eur. Phys. J. C \textbf{70} 635 (2010).%
%
\bibitem{Amsler:1995gf}
  C.~Amsler {\it et al.} [Crystal Barrel Collaboration],
  ``High statistics study of $f_0(1500)$ decay into $\pi^0\pi^0$,''
  Phys.\ Lett.\ B {\bf 342}, 433 (1995).
%
\bibitem{Amsler:1995bz}
  C.~Amsler {\it et al.} [Crystal Barrel Collaboration],
 ``High statistics study of $f_0(1500)$ decay into $\eta\eta$,''
  Phys.\ Lett.\ B {\bf 353}, 571 (1995).
%
\bibitem{Abele:1996nn}
  A.~Abele {\it et al.} [Crystal Barrel Collaboration],
  ``Observation of $f_0(1500)$ decay into $K_L K_L$,''
  Phys.\ Lett.\ B {\bf 385}, 425 (1996).
%
\bibitem{Amsler:2003bq}
C.~Amsler \textit{et al.} [Crystal Barrel Collaboration],
``Annihilation at rest of antiprotons and protons into neutral particles,''
Nucl. Phys. A \textbf{720}, 357 (2003).
%
\bibitem{Abele:1999tf}
A.~Abele \textit{et al.} [Crystal Barrel Collaboration],
``Evidence for a $\pi \eta$ P-wave in $\bar p p$ annihilations at rest into $\pi^0\pi^0\eta$,''
Phys. Lett. B \textbf{446}, 349 (1999).
%
\bibitem{Amsler:1994pz}
C.~Amsler \textit{et al.} [Crystal Barrel Collaboration],
``Observation of a new $I^G (J^{PC}) = 1^-(O^{++})$ resonance at 1450\,MeV,''
Phys. Lett. B \textbf{333}, 277 (1994).
%
\bibitem{Abele:1997qy}
A.~Abele \textit{et al.} [Crystal Barrel Collaboration],
``High mass $\rho$ meson states from $\bar p d$ annihilation at rest into $\pi^-\pi^0\pi^0$ spectator,''
Phys. Lett. B \textbf{391}, 191 (1997).
%
\bibitem{Abele:1999en}
A.~Abele \textit{et al.} [Crystal Barrel Collaboration],
``Anti-proton proton annihilation at rest into $K^+ K^-\pi^0$,''
Phys. Lett. B \textbf{468}, 178 (1999).
%
\bibitem{Wittmack:diss}
K.~Wittmack,
``Messung der Reaktionen $\bar p n\to K_SK^-\pi^0$ and $\bar p n\to K_SK_S\pi^-$'',
PhD thesis, Bonn (2001).
%
\bibitem{Abele:1998qd}
A.~Abele \textit{et al.} [Crystal Barrel Collaboration],
$\bar pp$ annihilation at rest into $K_L K^\pm \pi^\mp$
Phys. Rev. D \textbf{57}, 3860 (1998).
%
\bibitem{Abele:1999ac}
A.~Abele \textit{et al.} [Crystal Barrel Collaboration],
``The $\rho$ mass, width and line-shape in $p\bar p$ annihilation at rest into $\pi^+ \pi^- \pi^0$,''
Phys. Lett. B \textbf{469}, 270 (1999).
\end{thebibliography}
\end{document}